\documentstyle[prd,aps,floats,psfig]{revtex}
\begin{document}
\draft
%
%
\input epsf
\renewcommand{\topfraction}{0.8}
\twocolumn[\hsize\textwidth\columnwidth\hsize\csname
@twocolumnfalse\endcsname
\preprint{CIEMAT-00/XX, FTUAM-00/24}
\title{Measuring the Top Yukawa coupling to a heavy Higgs boson \\[1mm] 
at future {\boldmath $e^+ e^-$} Linear Colliders} 
\author{Juan Alcaraz}
\address{CIEMAT, Av. Complutense 22, 28040-Madrid, Spain}
\author{Ester Ruiz Morales} 
\address{Departamento de F\'{\i}sica Te\'orica, Universidad Aut\'onoma
de Madrid, Cantoblanco 28049-Madrid, Spain}

\date{\today}  \maketitle

\begin{abstract}
The determination of the Yukawa coupling of the top quark to the Higgs
boson is one of the most important measurements that a future $e^+e^-$
linear collider could provide. For a Higgs boson of mass greater than
350 GeV, this coupling can be determined using the Higgs resonant
contribution to $t \bar t$ production from $W^+ W^-$ fusion  at TeV
energies. We have made a careful evaluation of the  significance with
which the signal of a Higgs decaying to $t \bar t$ pairs  could be
observed at future $e^+ e^-$ linear colliders, with center  of mass
energies close to 1 TeV and an integrated luminosity of 1~ab$^{-1}$.
We find that a signal significance greater than 5 $\sigma$ and a
relative error in the top Yukawa measurement better than 10\%  can be
achieved, for Higgs masses in the $350-500$ GeV and $350-650$ GeV
ranges at facilities with 800 GeV and 1 TeV energies respectively.
\end{abstract}

\pacs{PACS numbers: 14.80.Bn, 14.65.Ha, 12.60.-i, 12.60.Fr
\hspace{2.5cm} Preprint: FTUAM-00/24}

\vskip2pc]

One of the most important tasks of future particle collider
experiments is to investigate the spontaneous breaking of electroweak
symmetry and, in particular, to understand the mechanism that
generates the masses of elementary particles. In the Standard Model
(SM) of electroweak interactions, fermion masses are generated
through the Yukawa interactions that couple the fermions to the Higgs
field. This mechanism implies that, after spontaneous symmetry
breaking, the fermion mass $m_f$ and its coupling $y_t$ to the physical
Higgs boson  are related by $m_f = y_t  v$, where $v$ is the
Higgs vacuum expectation value. The experimental verification of this
relation by the independent measurement of the fermion masses and
their Yukawa couplings provides an essential test of the fermion mass
generation mechanism of the SM.

This test has particular theoretical interest in the case of the top
quark. Since the mass of the top quark is close to the electroweak
scale,  it must have couplings of order one to the symmetry breaking
sector.  Thus it is expected to be a very sensitive probe of possible
new physics that might be responsible for   generating the top quark
mass. For example, alternative models of electroweak symmetry breaking
with new strong  interactions, like Technicolor and Topcolor models,
substantially modify the top quark interaction with the Higgs  sector
and give rise to new signals that could be studied at  future $e^+ e^-$ 
colliders \cite{RMP}. In spite of the interest of these new
physics signals, we will concentrate here on a  study within the SM,
given the intrinsic importance of the top  Yukawa coupling.

Experimentally, the determination of the top quark Yukawa coupling is a
challenging measurement that requires future accelerators with the
highest center of mass (CM) energies and luminosities. If the Higgs boson
mass $m_H$ is below 120 GeV, the top Yukawa coupling could be measured
in associated $t\bar t H$ production, with a statistical error of the order
of 10\% both at the CERN LHC\cite{AtlasTDR} and at high energy  
$e^+ e^-$ linear colliders (LC)\cite{BDR,JM}. 
But associated production is no longer efficient for higher Higgs masses.
For a Higgs boson heavier than 350 GeV,
the top Yukawa coupling could be determined from the Higgs decay 
into $t \bar t$ pairs.
This decay can be studied in high energy $e^+ e^-$ colliders
using the Higgs resonant contribution to the process
$W^+ W^- \to t \bar t$\cite{BT}. Unfortunately, this electroweak 
$t\bar t$ production process cannot be
observed at the LHC, due to the huge QCD background of $t \bar t$
production by gluon fusion. 

\begin{figure}[t]
\centering
\hspace*{-4mm}
\leavevmode\epsfysize=7cm \epsfbox{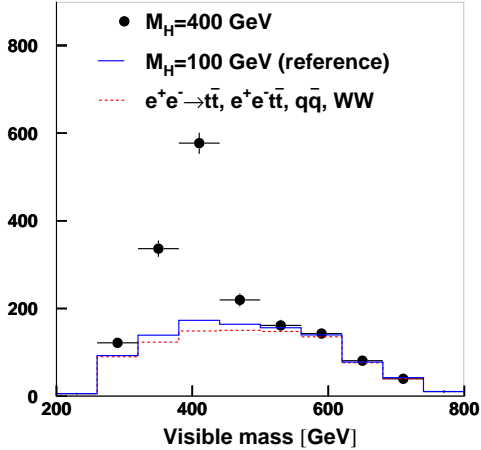}\\[3mm]
\caption[fig1]{\label{fig:H400}  Expected number of reconstructed 6-jet 
events as a function of the visible mass,
at $\sqrt s = 1 $ TeV and with ${\cal L}=$1 ab$^{-1}$ after all cuts,
for a Higgs boson of 400 GeV (dots) and for the backgrounds (dashed).
The expectations (including background) for a Higgs of 100 GeV 
(solid) are also shown for comparison. }
\end{figure}

\begin{figure}[t]
\centering
\hspace*{-4mm}
\leavevmode\epsfysize=7cm \epsfbox{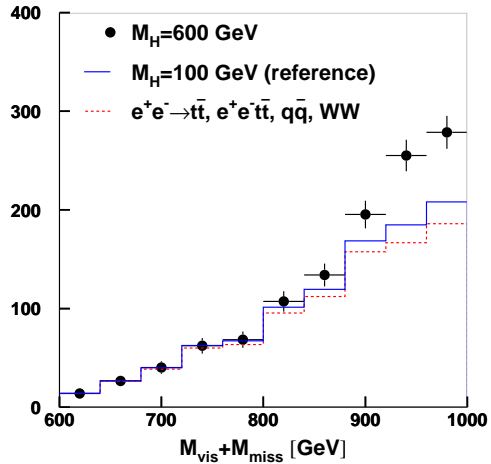}\\[3mm]
\caption[Fig2]{\label{fig:H600}  Sample of reconstructed 6 jet 
events as a function of the visible plus missing mass, 
at $\sqrt s = 1 $ TeV and with ${\cal L}=$1 ab$^{-1}$.
The event numbers correspond to  $m_H= $ 600 GeV (dots) 
and the sum of backgrounds (dashed). The expectations for the 
$m_H=$100 GeV case (solid) are also shown for comparison. }
\end{figure}

In this letter, we present the results of a complete simulation study
of the $ W^+ W^- \to H \to t \bar t$ process at future $e^+ e^-$ linear
colliders, including realistic backgrounds and experimental effects.
The aim is to obtain a reliable evaluation of the significance 
with which the signal of a Higgs boson decaying to a $t \bar t$ pair 
could be observed, and the accuracy that could be reached in the 
measurement of the top quark Yukawa coupling through this process 
at the planned accelerators. We have considered two designs 
with CM energies of 800 GeV and 1 TeV respectively, 
assuming an integrated luminosity of 1000 fb$^{-1}$.
Beamstrahlung has been simulated with the program CIRCE~\cite{CIRCE}, 
using the parameters of the TESLA linear collider\cite{TESLA}. 
The detector effects have been simulated with SIMDET~\cite{simdet}, 
a fast simulation program which reconstructs energy-flow objects according 
to the TESLA detector parameters. A similar performance 
is expected in other accelerator and detector designs, so our results
in the framework of the TESLA project also apply
to other $e^+e^-$ facilities operating at the same center-of-mass
energy and luminosity.

Some care was needed to make an accurate evaluation of 
the $ e^+ e^- \to t \bar t \nu \bar \nu$ signal, which includes 
as a subprocess the $WW$-fusion reaction $W^+ W^- \to t \bar t$. 
We first simulated the signal in the $WW$-fusion approximation
using the event generator Pandora\cite{pandora}. This
calculation was based on the effective-W approximation, using
the helicity and $p_T$-dependent $W$ structure functions of 
Ref. \cite{FW} and on-shell $WW \to t \bar t$ scattering amplitudes.
Comparing with a full SM calculation using the program CompHep~\cite{comphep} 
we have found that, at colliders with CM energies 
of 1.5 TeV or above, the improved $WW$-fusion calculation approximates well  
the exact result. However, at energies of 1 TeV or below, 
the $WW$-fusion approximation is not reliable, because there are
sizeable interference effects
between fusion and non-fusion diagrams that give the same
$t \bar t \nu \bar \nu$ final state. Since this effect cannot be neglected 
or taken into account by adding a reducible, non-interfering background,
an event generator including the full $e^+ e^- \to t \bar t \nu \bar \nu$ 
amplitudes is needed at the CM energies of our study.

To supply this, we have used the computer code NextCalibur~\cite{nextcalibur}, 
an upgraded version of the $e^+ e^- \to 4 f$ SM generator 
Excalibur. This new version has an improved simulation of the
initial state radiation (ISR) and includes 
the effects of the finite top mass and the Higgs exchange diagrams
which are crucial in our analysis.
We had to extend this program to include beamstrahlung effects and 
to have control on the helicity of the produced $ t \bar t$ pairs,
since the top and antitop decays into the final fermions strongly depend
on their polarization state. 
We then made the top and antitop quarks decay using the routines
of Pandora, which fully take into account spin correlations
in the $t \to b W \to b f f'$ decays
and include the finite width effects of the $W$ boson and the top quark.
The final hadronization of the decay products was
made with PYTHIA~\cite{pythia}.

\begin{table}
\begin{tabular}{ccccc} 
$m_H$ & $\sigma(LL) = \sigma(RR) $  & $\sigma(LR)$ 
 & $\sigma(RL)$ & Total \\\hline & & & & \\
100 & 0.11 & 0.22 & 0.21 & 0.65 \\
400 & 1.89 & 0.34 & 0.33 & 4.45 \\
600 & 0.89 & 0.22 & 0.21 & 2.21 \\
800 & 0.33 & 0.22 & 0.21 & 1.09\\[2mm] 
\end{tabular}
\vspace{3mm} 

\caption[Tab1]{\label{tab:sigmas} Helicity cross sections (in fb) for the 
$e^+ e^- \to t \bar t \nu \bar \nu$ process, 
at $\sqrt s = 1$ TeV and for different Higgs boson 
masses (in GeV). $\sigma(\lambda,\lambda^\prime)$ denotes the cross section 
for production of a top with helicity
$\lambda$ and an antitop with helicity $\lambda^\prime$. CP invariance implies
$\sigma(RR)=\sigma(LL)$.}
\end{table}

The helicity cross sections for the $e^+ e^- \to t \bar t \nu \bar \nu$ 
signal process, including ISR and 
beamstrahlung effects, are given in Table \ref{tab:sigmas},
for a collider CM energy of $\sqrt s = 1$ TeV and different Higgs boson 
masses. The total signal cross sections range 
from 0.65 fb for $m_H =$ 100 GeV up to
4.45 fb for $m_H= 400$ GeV where the largest sensitivity
to the Higgs boson contribution is achieved.
The signal events look very much like $t \bar t$ events, but with a
lower visible mass in general and a 
substantial missing longitudinal and transverse momenta 
(of order $M_W/2$) carried away by the two electron neutrinos. 
Due to the missing momenta in the longitudinal and transverse directions,
only the final 6-jet events in which both the top and antitop 
decay into a b quark plus two additional quarks can be fully reconstructed 
experimentally. For this reason, we restrict our analysis to
events with 6 jets in the final state.

The main backgrounds to this signal have been generated with PYTHIA.
There are huge backgrounds from $q \bar q$ and 
$W^+ W^-$ production, with total cross sections before cuts of 3400 fb and 
3700 fb respectively at $\sqrt s$= 1 TeV. Fortunately, they present 
event shapes very different to the signal, so that they are largely reduced 
by applying hard cuts.
We require no isolated leptons in the final state and force the event 
into six jets (Durham Algorithm) with $Y$  greater than 5$\times 10^{-4}$.
We have also imposed loose cuts in the event thrust, major and 
C-parameter to reject back-to-back events, and a moderate b-tagging
based on consistency with primary vertex. 

More dangerous backgrounds are direct $t \bar t$ production, with a cross
section of 243 fb and $e^+e^- t \bar t $ production with 17 fb.
This latter background comes mainly from $\gamma \gamma$ fusion and
can be reduced requiring the missing transverse energy in the event to be
greater than 50 GeV.
We have also required missing mass greater than 200 GeV,
to suppress $t\bar t$ production with ISR and to cut the $Z$ peak 
from $Z t \bar t$ production with the $Z$ decaying into two neutrinos.

After these cuts,
there is still  too large a contribution from 
misreconstructed $e^+ e^- \to t \bar t$ events.
This background can be efficiently reduced by choosing the jet 
association that gives the best fit to the reconstructed $t$ and $W$
masses and keeping events within five standard deviations of the 
expected values. 
After all cuts, the acceptances for the signal
events are in the range of 18\% to 12\% for $\sqrt s = 1$ TeV 
and Higgs masses from 400 GeV to 800 GeV. The 
$WW$ and $q \bar q$ backgrounds are largely reduced by a  
$5 \times 10^{-6}$ rejection factor, becoming a small
percentage of the background composition which is
dominated by the $e^+ e^- \to t \bar t$ and $e^+e^- \to e^+e^- t \bar t $ 
samples.

In order to evaluate the significance of 
the $W ^+ W ^- \to H \to t \bar t$ signal,
we have chosen as final reference background the data expectation 
in the case of a light Higgs boson ($m_H=100$ GeV). 
The signal to background ratio strongly
depends on the Higgs mass hypothesis. For $m_H = 400$ GeV, we expect
a total number of 695 reconstructed 6-jet events for the signal 
over 993 events of background. This gives a visible mass distribution
nicely peaked at the expected Higgs mass value, leading to a significant
observation as shown in Fig.\ref{fig:H400}. The situation gets worse as 
the Higgs mass increases, giving only 52 signal events at $m_H=800$ GeV.
For higher values of $m_H$, the most sensitive distribution is the sum
visible plus missing mass, which strongly peaks at values close to $\sqrt s$
when the Higgs boson is produced almost at rest.
This distribution is also shown in Fig.\ref{fig:H600}, in the case of 
$m_H =  600$ GeV.

\begin{figure}[t]
\centering
\hbox{\hspace{-0.6cm}\psfig{file=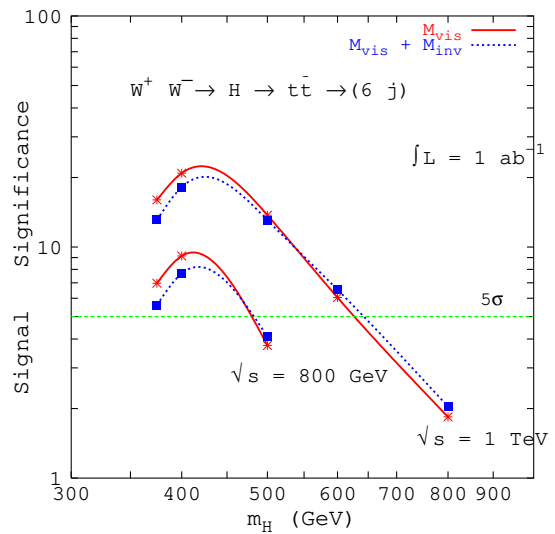,width=10.2cm,angle=-90}
\vspace{3mm}}
\caption[Fig3]{\label{fig:sign} Expected significance 
of the $H\to t \bar t$ 
signal as a function of the Higgs mass, from 6-jet events at TESLA. 
}
\end{figure}

\begin{figure}[h]
\centering
\hbox{\hspace{-0.6cm}\psfig{file=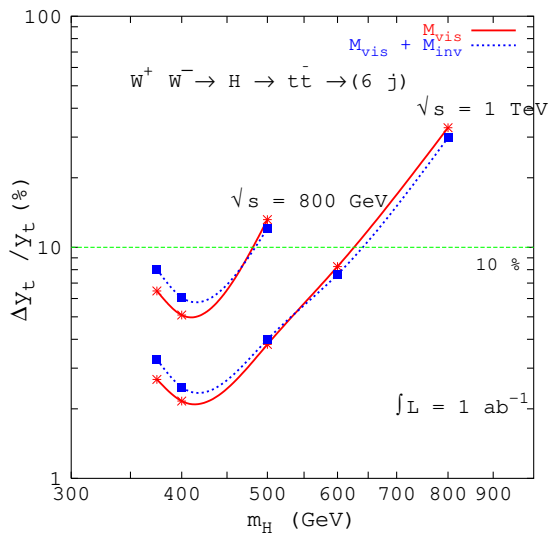,width=10.2cm,angle=-90}
\vspace{3mm}}
\caption[Fig4]{\label{fig:yuk} Expected relative precision in the top Yukawa 
measurement as a function of the Higgs mass, from 6-jet events at TESLA.}
\end{figure}

\begin{table}[t]
\vspace*{2mm}
\begin{tabular}{ccccc} 
&\multicolumn{2}{c}{$\sqrt s$ = 1 TeV } &
\multicolumn{2}{c}{ \hspace{3mm} $\sqrt s$ = 800 GeV } \\ \hline
$m_H$  & {Signif. ($\sigma$)} & {$\Delta y_t / y_t$(\%)} & 
{Signif. $(\sigma)$ }& 
{${\Delta y_t / y_t}$(\%)} \\
\hline & & & & \\
375 & 16  [13] & 2.7 [3.3] & 7.0 [5.6] & 6.5  [8.3]\\
400 & 21  [18] & 2.2  [2.5] & 9.1 [7.7] & 5.1 [6.1]\\
500 & 14  [13] & 3.8  [4.0] & 3.8 [4.1] & 13 [12] \\
600 & 6.0  [6.5] & 8.2 [7.6] &-- & -- \\
800 & 1.8  [2.0] & 33 [30] &  -- & -- \\[2mm] 
\end{tabular}
\vspace{3mm} 

\caption[Tab1]{\label{tab:all} Expected significance and relative precision
in the top Yukawa measurement for different Higgs masses at TESLA,
obtained from the binned fit to the visible mass distribution. 
The results from the fit to the visible plus missing 
mass distribution are also given in brackets.}
\end{table}

The results of the analysis are summarized in Figs. \ref{fig:sign} and
\ref{fig:yuk} and in Table \ref{tab:all}. The
significance of the signal is estimated by  binned fits to the visible 
and the visible plus missing mass distributions. The measured data in each 
bin, $N$, are assumed to follow the dependence $N = x~S + B$, where 
$S$ and $B$ are the signal and background expectations, respectively. The 
significance is given by $x / \delta x$, where $x$ is the fitted factor
(1 in the SM) and $\delta x$ is the statistical error from the fit. The 
results guarantee that at $\sqrt s= 1$ TeV a signal can 
be observed at the level of $5$ standard deviations if the
SM Higgs mass is in the $2 m_t < m_H < 650$ GeV range. Systematic effects 
are not expected to modify significantly this result.
The signal of a $H \to t \bar t$ decay could be established with 
more than 20 $\sigma$ if the Higgs boson mass is close to
400 GeV.
As expected, the visible mass distribution is more sensitive than
the visible plus missing mass distribution for lower Higgs mass values, 
and the situation is reversed at Higgs masses 
approaching $\sqrt s$. The analysis 
has been repeated at a lower CM energy of $\sqrt s=800$ GeV. In 
this case the 5 $\sigma$ range of observation is reduced to 
$2 m_t < m_H < 500$ GeV, due to a lower signal cross section and
a larger cross section for the $e^+ e^- \to t \bar t$ background.

The significance of the signal can be reinterpreted as a direct measurement 
of the $Ht \bar t$ Yukawa coupling $y_t$. These results are 
presented in Fig. \ref{fig:yuk} and in Table \ref{tab:all}. 
The main conclusion is that, for 
all cases in which there is a 5 $\sigma$ evidence for the signal, the Yukawa 
coupling can be measured with a relative precision better than $10\%$. 
A maximum accuracy close to 2\% can be achieved for Higgs masses 
around 400 GeV. It is also remarkable that it would be feasible to establish a 
non-vanishing Yukawa coupling at the $95\%$ confidence level for masses as 
high as $m_H=800$ GeV at a $\sqrt s= 1$ TeV collider, provided that
the possible sources of systematic uncertainties are kept under control.

To conclude, we have shown that if the Higgs boson is heavier than 350 GeV,
it will be possible to establish the signal of the Higgs decaying 
to top-antitop pairs at future high energy $e^+ e^-$ colliders, giving a 
good determination of the value of the top-Higgs Yukawa coupling. 
This would be an important measurement that cannot be done at the LHC. 
The significance of the signal is impressive for a $\sqrt s$= 1 TeV collider,
and reasonably good at $\sqrt s $=800 GeV. It should be remembered that, 
if the Higgs boson mass is in the $350-600$ GeV range, it will be seen 
first at the LHC through its decay into vector bosons, 
with significances of the order of $40-30$ standard deviations \cite{AtlasTDR} 
assuming full luminosity of $100$ fb$^{-1}$. The possibility of
observing the Higgs decay into $t \bar t$ pairs with a similar significance, 
as given in Table \ref{tab:all}, is a remarkable and largely unexpected
result of our analysis.

The signal significance could be substantially increased using tighter 
cuts and more sophisticated methods in the event and data analysis. 
Several tools that depend crucially on an optimal detector performance,
like kinematic fitting, a high degree of b or c-tagging, polarization 
analysis in top decays, etc..., have not been used in this study.
This gives our analysis a high degree of robustness, and allows 
these results, obtained in the framework of the 
TESLA project, to be safely extrapolated to other 
$e^+e^-$ accelerator and detector designs.
Moreover, we can expect a similar degree of signal observability
in other physical scenarios, like Technicolor and Topcolor models, 
in which new resonances from the symmetry breaking sector give 
effects of similar strength as the SM Higgs boson contribution.

ERM is grateful to Michael Peskin for his encouragement and advice
in the study of the $W^+W^-\to t \bar t$ process over the past years.
We also thank him and J. F. de Troc\'oniz for their careful reading of
the manuscript and for useful comments. This work has been supported by the 
Spanish CICYT grants AEN99-0305 and AEN97-1768.

\end{document}